\documentclass[useAMS,usenatbib]{mn2e}               
\usepackage{graphicx}

\def\farcs{\hbox{$.\!\!^{\prime\prime}$}}
\def\sun{\hbox{$\odot$}}
\title[CO observations of restarting radio galaxies]{A search for molecular
gas in restarting radio galaxies}
\author[Saripalli and Mack]
  {L.~Saripalli,$^{1,3}$\thanks{E-mail:lsaripal@rri.res.in}
  K.-H.~Mack,$^2$\thanks{E-mail:mack@ira.inaf.it} \\
  $^1$CSIRO ATNF, Locked Bag 194, Narrabri, NSW 2390, Australia\\
  $^2$INAF-Istituto di Radioastronomia, Via P. Gobetti 101, I-40129
       Bologna, Italy\\
  $^3$Raman Research Institute, C. V. Raman Avenue, Sadashivanagar, 
   Bangalore, India}

\begin{document}

\label{firstpage}

\maketitle

\begin{abstract}

Several radio galaxies are known that show radio morphological signatures that are best
interpreted as restarting of nuclear activity after a period of quiescence. The conditions
surrounding the phenomenon of nuclear recurrence are not understood. In this paper we
have attempted to address this question by examining the nuclear fuelling characteristics
in a sample of restarting radio galaxies. We have examined the detection rate for molecular 
gas in a representative sample of nine restarting radio galaxies, for seven of which we
present new upper limits to the molecular gas mass derived from CO line observations we made
with the IRAM 30-m telescope. We derive a low CO detection rate for the relatively young 
restarted radio galaxies suggesting that the cessation of the nuclear activity and its
subsequent restarting may be a result of instabilities in the fuelling process rather than
a case of depletion of fuel followed by a recent fuel acquisition. It appears that abundant 
molecular gas content at the level of few $~10^{8}$ -- $ 10^{9}~M_{\sun}$ does not necessarily
accompany the nuclear restarting phenomenon.
For comparison we also discuss the molecular gas properties of five normal
giant radio galaxies, three of which we observed using SEST. Despite
obvious signs of interactions and nuclear dust disks none of them has been
found to host significant quantities of molecular gas.

\end{abstract}

\section{Introduction}
One of the exciting developments in radio galaxy research in recent years 
has been 
the recognition of a restarting of nuclear activity following a quiescent
period. The usual extended twin-lobe radio morphology of radio galaxies are
in some cases found to be accompanied by a second pair of lobes sharing the
same central radio core (Roettiger et al. 1994; Subrahmanyan, Saripalli \& Hunstead 1986; 
Schoenmakers et al. 2000a). The `double-double' structures were interpreted by these 
authors as radio galaxies that were experiencing a second epoch of activity. 
Detailed studies of some of the more striking examples of this class of 
radio galaxies (e.g. Lara et al. 1999; Schoenmakers et al. 2000b;
Saripalli, Subrahmanyan 
\& Udayashankar 2002; Saripalli, Subrahmanyan \& Udayashankar 2003) have 
estimated the ages of the second activity epoch to be few times $10^{6}$ yr 
suggesting that the restarted activity is fairly recent in 
comparison with the ages of radio galaxies in general (Scheuer 1995; Mack et al. 1998). 

The nuclear activity that manifests itself as extended twin lobe radio 
structures in radio galaxies is believed to be the result of gas being 
accreted on to a super-massive black hole located at the centre. There is
evidence that gas at the centres of active galaxies forms an accretion disk
via which it gets transported to the vicinity of the
Schwarzschild radius of a few thousandths of a parsec. An important component 
of the central region of an active galaxy is a torus of dust and gas 
situated at much larger scales from several hundred parsec to a few kiloparsec.
Observations have targeted this region for clues to the presence
of a fuel reservoir powering the active galaxies (e.g. Mazzarella et al. 1993; 
Lim et al. 2000). The fuel searched for at millimetre 
wavelengths is the CO gas since next to molecular hydrogen, that has only
weak emission lines, the CO molecules are the most abundant. 

The causes behind the restarting of nuclear activity are far from clear. 
Suggestions have been speculative and have included `long term variability'
in central engine output (Roettiger et al. 1994) and accretion disk 
instabilities (e.g. Schoenmakers et al. 2000a). 
In this paper we explore this question of why radio galaxies become inactive
and then restart their nuclear activity after several million years. We look
at the causes from the point of view of fuel availability to the central
engine. 
Starting from the basic premise
that all AGN require fuel for their activity we enquire whether the inactivity
is caused by an interruption to the fuelling process or a depletion of the fuel
store that sustained the activity. Are the manifestations 
we see (of nested, inner double lobe structures) a result of a new
accumulation of fuel after its depletion in the earlier epoch or a
continuation of the old fuelling process after an interruption? 
\begin{table}
  \caption{The sample}
\begin{tabular}{|lllrcc}
\hline
Source  & RA (J2000) & Dec (J2000) & \multicolumn{1}{c}{z} & lin. size \\   
 &  [$^{\rm h}:\;\;^{\rm m}:\;\;^{\rm s}$] & [$^{\circ}:\;\;^{\arcmin}:\;\;^{\arcsec}]$ &  & [kpc] \\
\hline
4C\,12.03 & 00:09:52.6 & +12:44:05   & 0.156     &  577      \\  
4C\,29.30 & 08:40:02.4 & +29:49:03   & 0.065     &  97       \\
3C\,219   & 09:21:08.6 & +45:38:57   & 0.174     &  465      \\
3C\,236   & 10:06:01.7 & +34:54:10   & 0.101     & 4500      \\
4C\,26.35 & 11:58:20.1 & +26:21:12   & 0.112     & 514       \\
3C\,293   & 13:52:17.8 & +31:26:46   & 0.045     & 218       \\
3C\,388   & 18:44:02.4 & +43:33:30   & 0.092     & 82        \\
3C\,424   & 20:48:12.0 & +07:01:18   & 0.127     & 76        \\
3C\,445   & 22:23:49.6 & $-$02:06:12 & 0.056     &  596      \\
          &            &             &           &           \\
NGC\,315& 00:57:48.9 & +30:21:09     & 0.016     & 1080      \\
0319-454& 03:20:57.6 & $-$45:15:10   & 0.063     & 1789      \\
0503-286& 05:05:49.3 & $-$28:35:20   & 0.038     & 1800      \\
0511-305& 05:13:31.9 & $-$30:28:50   & 0.058     & 740      \\
NGC\,6251 & 16:32:32.0 & +82:32:16   & 0.025     & 1857      \\

\hline
\end{tabular}
\end{table}

The CO observations of powerful radio galaxies
made using the J(1--0) and J(2--1) line transitions have revealed interesting
results -- it is found that compact radio galaxies and edge darkened (FR-I) radio galaxies
are more frequently detected than larger radio galaxies with edge brightened (FR-II)
structures even considering their larger distances (given their sufficiently low upper limits; 
Evans et al. 2005 and references therein). Together with the high
detection rate for the ultra-luminous infra-red galaxies (ULIRGs) the gradation 
in CO detection rates prompted 
suggestions of an evolutionary link between the ULIRGs, compact radio sources and larger
radio galaxies (e.g. Mazzarella et al. 1993; Sanders et al. 1988). The underlying hypothesis 
was that as the radio galaxies form and evolve there is a depletion of fuel as it is
processed. In this sequence the ULIRGs are at initial stages of fuel accumulation 
and the compact radio sources, which form after nuclear activity is triggered, 
are in the early phase of nuclear activity with the large FR-II being at later 
stages. The large FR-Is, with their higher detection rate 
of molecular gas, do not fit into this picture.

The notion of consumption of fuel as the source grows has implications for the 
formation of the `double-double' radio sources. With the restarting radio galaxies 
found to be relatively young, they should be more amenable to the detection of molecular
gas than larger radio galaxies if the new activity resulted from a recent fuel
accumulation.

We have compiled a sample of radio galaxies considered to be sources with
restarted nuclear activity with the aim of exploring their fuel aspects. We
expect that the CO detection rate among this class of radio galaxies is 
intermediate between the detection rate of compact radio sources and larger
radio galaxies if they have restarted due to a recent acquisition of fuel. If
however the new activity is a consequence of an interruption to the 
original fuelling process and subsequent resumption the sources should have
low or zero detection rate just as is the case for the large FR-II radio galaxies.
In this paper we present CO~J(1--0) and CO~J(2--1) observations of a small
sample of restarting radio galaxies made using the IRAM radio telescope. 

In a separate program, we observed three megaparsec-size
radio galaxies with the 15-m SEST in the 
CO~J(1--0) and CO~J(2--1) line transitions. We have included these data in the 
present paper for comparison (a) because they are radio galaxies at the extreme end of the
linear size distribution and hence among the oldest radio galaxies, (b)
they form a good contrast in age with respect to the 
radio galaxies with restarting nuclear activity and (c) following the model
proposed by Kaiser et al. (2000) they form the parent sample of most
restarting sources.

In the following section we present our sample. The observations are described
in Section 3. In Section 4 we give the results and this is followed with a 
discussion in Section 5. We adopt a flat cosmology with Hubble 
constant $H_{0}$ = 71~km~s$^{-1}$~Mpc$^{-1}$ and matter density parameter 
$\Omega_{m}$ = 0.27.

We have included from the literature four radio galaxies (NGC\,315, 
4C\,29.30, 3C\,293 and NGC\,6251) that were observed in the CO~J(1--0) and
CO~J(2--1) line transitions.
Two are normal giant radio galaxies (NGC\,315, NGC\,6251) that do not show
evidence for 
restarting of nuclear activity and two, 4C\,29.30 and 3C\,293 exhibit strong
evidence for a recurrence
in their nuclear activity (Jamrozy et al. 2006, submitted; Akujor et al. 1996).
The discussion
section will be based on the 9 restarting radio galaxies and 5 normal giant
radio galaxies.

\section{Sample selection}

In Table~1 we list the two samples, the 9 restarting radio galaxies and 
5 normal, giant radio galaxies. Both samples are representative samples of
their class of sources, neither being a complete sample.
The restarting radio galaxy sample is part of a larger compilation of
sources from the literature that
have been recognised as having undergone a second epoch of nuclear activity.
Signatures of restarted activity are most often in the form of morphological 
structures such as an inner pair of bounded emission peaks within the extended
radio
lobes. 4C\,12.03 (Leahy \& Perley 1991), 4C\,26.35 (Owen \& Ledlow 1997), 3C\,219 
(Bridle et al. 1986), 3C\,424 (Black et al. 1992) and 3C\,445 (Leahy et al. 1997)
 all have bounded emission peaks within their
lobes. 3C\,236 and 3C\,293 are found to have such a pair of emission peaks well within 
their central compact cores (Schilizzi et al. 2001; Akujor et al. 1996). In one source,
3C\,388, there is 
radio spectral evidence for a second activity epoch -- the source is found to 
have radio lobes of a typical spectral index distribution embedded within two
outer features of much steeper spectral index values (Roettiger et al. 1994). 
In selecting the seven restarting radio galaxies for our observations we were
constrained by the
declinations accessible to the IRAM telescope and the redshifts of 
the sources. With the standard setup (as used by us) the IRAM observing band
allowed a redshift range of up to 0.42 and 0.17 for the
two CO~J(1--0) and J(2--1) frequencies respectively. 

The sources observed with the SEST were all large sized radio galaxies with
linear sizes in the megaparsec range. Three radio galaxies (0319$-$454,
0503$-$286 
and 0511$-$305) were selected with
declinations suitable for observations with SEST. The sources had redshifts
that allowed CO~J(1--0) and J(2--1) line transitions to be observable within
the SEST receiver bandwidths. 

The sources added from the literature were observed with different telescopes:
NGC\,315 with the IRAM 30-m telescope (Braine et al. 1997), 4C\,29.30 and 3C\,293 with
the NRAO 12-m telescope (Evans et al. 2005) and NGC\,6251 with the OSO 20-m telescope
(Elfhag et al. 1996).

\begin{figure*}
\hspace*{-0.5cm}
\resizebox{8.95cm}{!}{\includegraphics{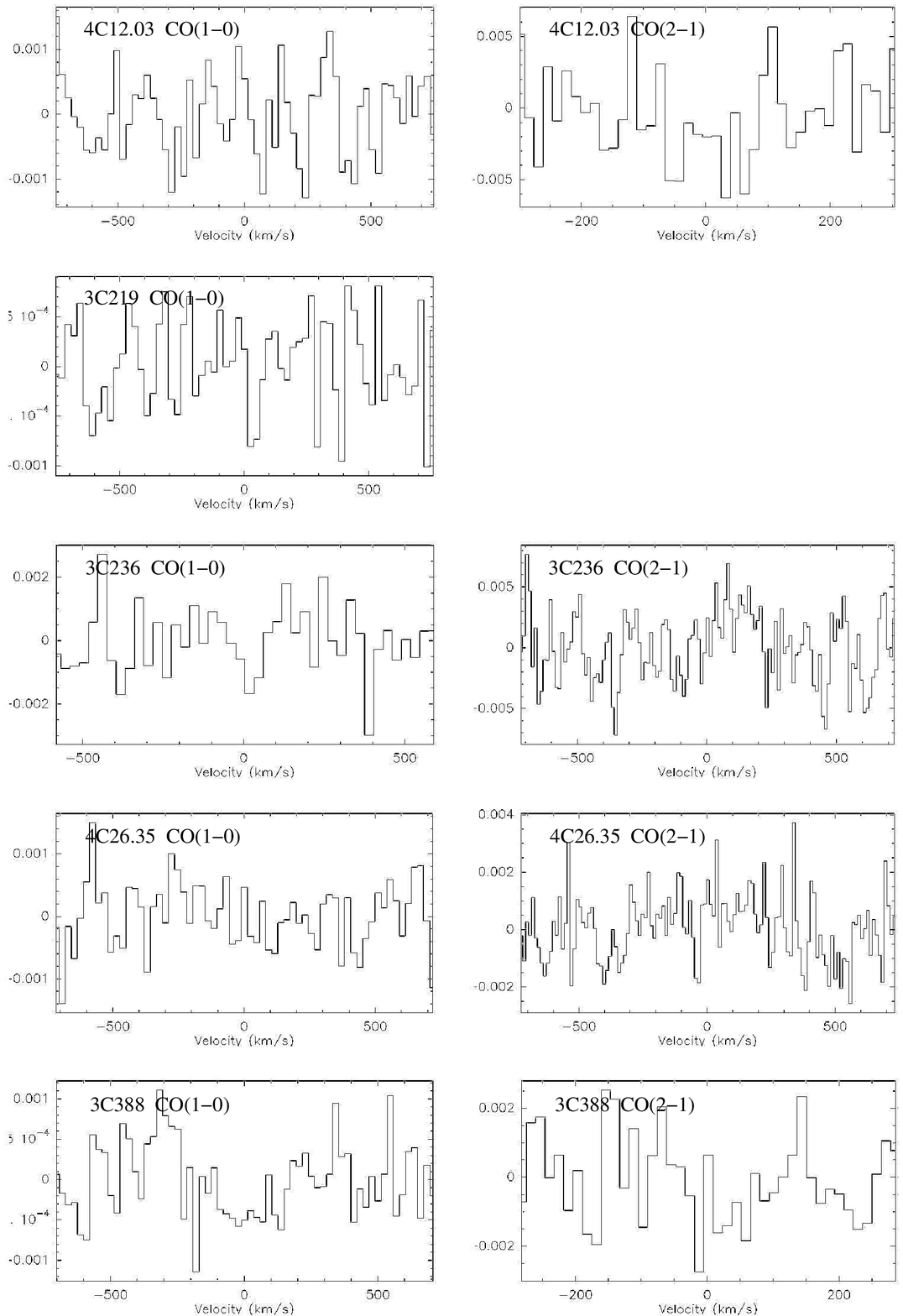}}
\hfill
\resizebox{9.05cm}{!}{\includegraphics{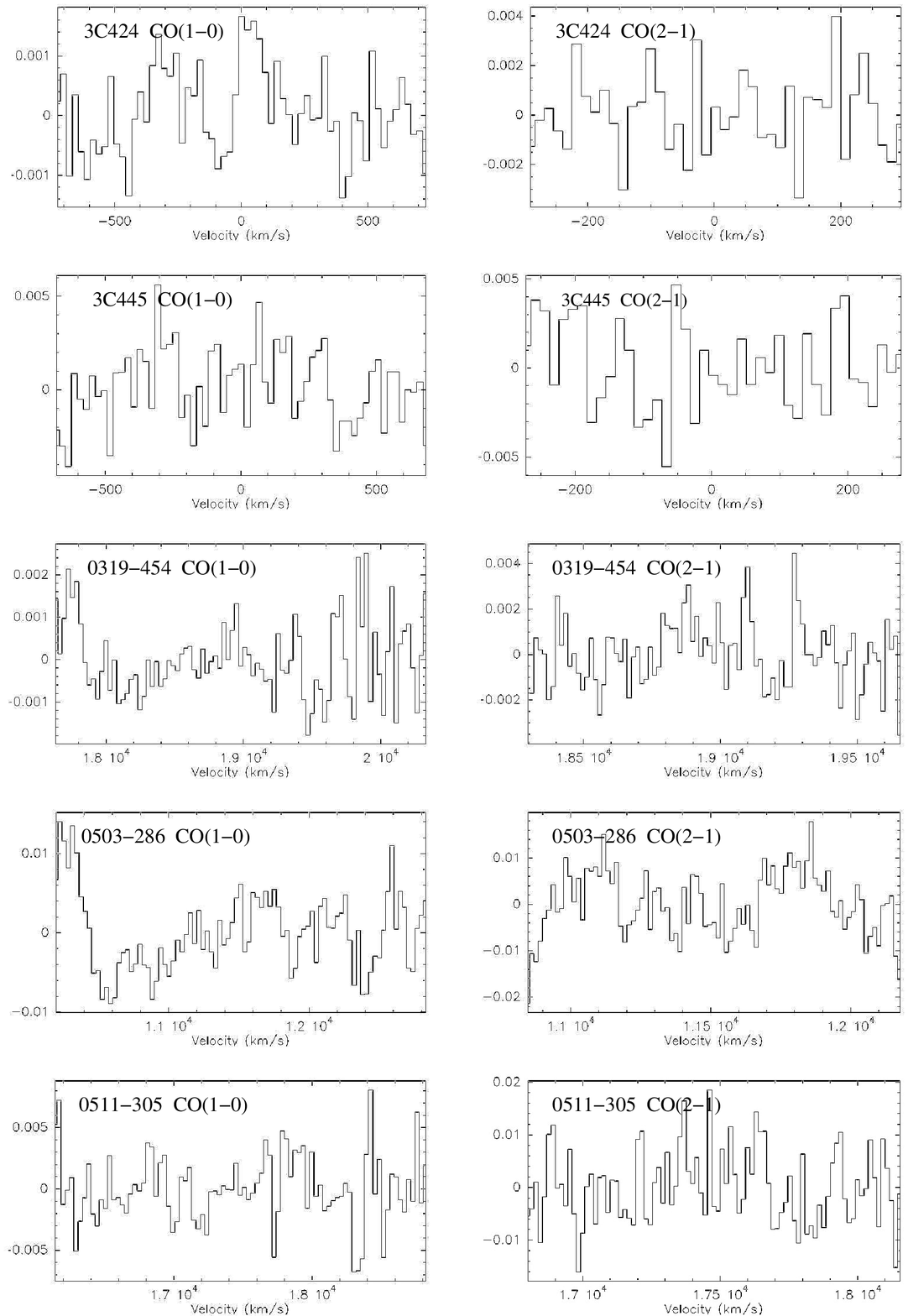}}
\caption{The CO spectra of our target sources. The CO(2-1) spectrum of 3C\,219 could not
be obtained since the A230/B230 receivers of the 30-m telescope could not be tuned to the required
redshifted frequency. The temperature scales of the ordinates are calibrated in Kelvin on the
${\rm T}^*_{A}$ scale.}
\end{figure*}

\section{Observations and data analysis}

\subsection{IRAM observations}
At the IRAM 30-m telescope we used the A100/B100 receiver combination for the
CO(1-0) line and the A230/B230 combination for the CO(2-1) line search.
The beam-sizes were $23\farcs4$ 
and 11\farcs7 (at the typical frequency of 105 GHz and 210 GHz respectively).
The observations were mostly performed in wobbler secondary mode with a 
beam throw between on and two symmetric off-source positions of 240$''$ 
and a switching rate of 0.5 Hz.
The observation of 3C\,445 and about two thirds of the 4C\,12.03 scans had
to be accomplished in a symmetrical position switching mode with offsets of
$-80''$ and $+80''$ from the source position.  
All signals were connected to the VESPA autocorrelator with bandwidths of 640
MHz and spectral resolutions of 1.25 MHz. In addition and for redundancy,
the A100 and B100 receivers were also linked to the 1-MHz filterbank with
512 MHz bandwidth for each part and a spectral resolution of 1 MHz. The observations 
were spread over three separate terms. The last term (June 2006) when sources
3C\,236, 3C\,219 and 4C\,26.35 were observed was carried out in the presence of 
occasional wind gusts.
During this session the A230/B230 signals were connected to the 4-MHz filterbank
instead of the VESPA autocorrelator which provides a total bandwidth of 1024 MHz.
More detailed observational parameters are compiled in Table~2.

\subsection{SEST observations}
The observations were carried out using the 15-m Swedish-ESO Millimetre
Telescope (SEST) in December 2001. The IRAM 115/230~GHz receiver
was used in combination with two low-resolution spectrometers that afforded 
bandwidth of 500~MHz at 105~GHz and 1000~MHz at 210~GHz and beam-sizes of 49\farcs4 and 25\farcs2,
respectively. The source spectra were obtained 
in the dual beam-switching mode using a chopper wheel with two symmetric reference positions 
separated by $11'$ and $50''$ in azimuth. 
For calibration the chopper wheel method was employed and observations of a
hot and cold load were made at regular intervals. Telescope pointing checks were carried
out regularly. 
 
Unfortunately as much as 60\% of the allocated time was lost due to poor weather. High
humidity was a concern for a significant length of the remaining time.\\

\subsection{Data analysis}
The data was analysed using the CLASS software. For both sets of data we examined the 
spectral data on each of the sources for each of the CO transitions. We rejected scans that
were affected by system failures and scans that had ripples in the baselines.
Although summer observing conditions 
are not particularly favourable for 1-mm observations we nevertheless recorded CO~J(2--1) line 
emission simultaneously with the CO~J(1--0) frequency.

For each of the spectra the continuum was fit with a first-order polynomial baseline
after dis-regarding the channels at the edge of the band.
The rms noise was estimated after subtracting the continuum.

The spectra were smoothed to reach a final velocity resolution of about 22 km$\,$s$^{-1}$ 
for CO(1-0) and about 14 km$\,$s$^{-1}$ for CO(2-1), for the exact values we refer to Table 2.
Throughout the article, unless explicitly stated otherwise, the brightness temperatures are expressed in the
${\rm T}_{\rm mb} = (\eta_{\rm f}/\eta_{\rm mb}){\rm T}^*_{A}$ scale, where $\eta_{\rm f}$ is the forward
efficiency (Pico Veleta: 0.95 and 0.91 at 105 and 210 GHz, SEST assumed to be unity) and
$\eta_{\rm mb}$ is the main beam efficiency (Pico Veleta: 0.75 and 0.57 at 105 and 210 GHz; 
SEST: 0.70 and 0.50 at 105 and 210 GHz).
\begin{table*}
  \caption{Observing Log}
\begin{tabular}{|lcrrccccccc|}
\hline
Source  & line transition  &  time & frequency & $\sigma_{\rm mb}$ & vel. res.  &date&telescope & $\tau$  & T$_{\rm sys}$  \\   
 &  &  \multicolumn{1}{c}{[min]} &  \multicolumn{1}{c}{[MHz]}    & [mK] & [km$\;s^{-1}$] & &    &     
             & [K]	    \\ 
\hline
4C\,12.03 & CO(1-0) &    158&    99715.574&   0.85 & 24.05  &  Jul 2003&    PV& 0.06  & 121\\   
          & CO(2-1) &    158&   199427.327&   4.39 & 15.03  &  Jul 2003&    PV& 0.33  & 410\\ 
3C\,219   & CO(1-0) &    222&    98151.769&   0.55 & 24.44  &  Jul 2006&    PV& 0.04  & 120\\  
          &    ---  &       &             &        &   &          &      &       &    \\
3C\,236   & CO(1-0) &    111&   104906.447&   1.39 & 22.86  &  Jun 2006&   PV & 0.11  & 147\\
          & CO(2-1) &    111&   209808.873&   4.16 & 11.43  &  Jun 2006&   PV & 0.42  & 449\\
4C\,26.35 & CO(1-0) &    310&   103661.155&   0.63 & 23.14  &  Jul 2006&   PV & 0.06  & 130\\
          & CO(2-1) &    310&   207318.336&   1.81 & 11.57  &  Jul 2006&   PV & 0.24  & 298\\
3C\,388   & CO(1-0) &    452&   105656.466&   0.57 & 22.70  &  Sep 2004&   PV & 0.11  & 132\\ 
          & CO(2-1) &    452&   211308.882&   2.00 & 14.19  &  Sep 2004&   PV & 0.32  & 307\\ 
3C\,424   & CO(1-0) &    281&   102282.366&   0.84 & 23.45  &  Jul 2003&   PV & 0.09  & 155\\ 
          & CO(2-1) &    280&   204560.812&   2.27 & 14.66  &  Jul 2003&   PV & 0.36  & 532\\ 
3C\,445   & CO(1-0) &     56&   109137.667&   2.83 & 21.98  &  Jul 2003&   PV & 0.09  & 189\\ 
          & CO(2-1) &     42&   218272.151&   3.69 & 13.73  &  Jul 2003&   PV & 0.21  & 264\\ 
          &         &       &             &        &   &          &      &       &    \\
0319-454  & CO(1-0) &    273&   108408.866&   0.96 & 30.60  &  Dec 2001&  SEST& 0.07  & 214\\ 
          & CO(2-1) &    211&   216813.484&   2.04 & 15.47  &  Dec 2001&  SEST& 0.15  & 522\\ 
0503-286  & CO(1-0) &     24&   111008.456&   7.45 & 29.88  &  Dec 2001&  SEST& 0.20  & 300\\ 
          & CO(2-1) &     30&   222012.657&  14.11 & 15.10  &  Dec 2001&  SEST& 0.63  & 668\\ 
0511-305  & CO(1-0) &     46&   108921.020&   3.66 & 30.45  &  Dec 2001&  SEST& 0.23  & 325\\ 
          & CO(2-1) &     46&   217837.866&  14.00 & 15.39  &  Dec 2001&  SEST& 1.17  & 886\\ 
\hline
\hline
\end{tabular}
\end{table*}

\section{Results}
Our results for all the sources observed are presented in Table~2. In Col. 5 we give the rms value
of the observed main beam temperatures after smoothing to the velocity resolution given in Col. 6.
None of the sources was detected above the 3-$\sigma$ level.  
In order to derive upper limits to the molecular gas estimates for our sample of radio
galaxies we have used the expression given by Braine et al. (2000):
$$
M_{\rm H_2} = 2 I_{\rm CO}  m_{\rm p}  X \Omega D^2
$$ 
where $I_{\rm CO} = T_{mb} \Delta V$ for detected CO emission and 
$I_{\rm CO} = \sigma_{\rm mb} \sqrt{\Delta V \delta v}$ for upper limits.

Here $\sigma_{\rm mb}$ is the noise level derived from the spectra (in case of no detections, the
masses were calculated for 3-$\sigma_{\rm mb}$ values of the main beam temperature), $\Delta V$ the line
width (for the limits assumed to be 300 km$\,$s$^{-1}$), $\delta v$ the velocity resolution
(in km$\,$s$^{-1}$) and
$m_{\rm p}$ the proton mass. X represents the CO-to-H$_2$ conversion factor 
(no correction has been made for the presence of helium),
$\Omega$ corresponds to the beamsize (assumed Gaussian) and $D$ is the luminosity distance to the target.

To allow for a deviation of the X-factor for molecular clouds in our sample of galaxies 
from the value determined for our Galaxy we have used a range ($3.7 \cdot 10^{23}$ to $2.1 \cdot 
10^{24} {\rm m^{-2}}({\rm K\;km\; s^{-1})^{-1}}$).  
The lower value is determined for the ULIRGs (Solomon \& Vanden Bout 2005) whereas the higher 
value is appropriate for normal galaxies (Braine et al. 2000). Following de Blok \& van der 
Hulst (1998) we used a conversion factor for mass estimates based on the CO~J(2--1) transition 
which is 20\% higher than that based on CO~J(1--0) transition.

These values have been used to calculate the molecular
gas masses in Table~3 in units of $10^{9}~M_{\sun}$. The two columns are for
the two extreme values of the CO-to-H$_2$ conversion factor, as given above. As restarting 
radio galaxies should possess characteristics that are closer to the ULIRGs rather than
normal spirals we take the lower value to be a more realistic representation of the conversion factor
for our sources. 

\section{Discussion}

The work reported in this paper centres around radio galaxies that show strong evidence 
for a new epoch of activity at their centres. The new activity epoch manifests itself as
bounded, inner emission peaks within the extended radio lobes, in most cases seen as an embedded
pair. From Table~3 we see that only one out of nine restarting radio galaxies 
is detected in CO. The non-detections translate to 3-$\sigma$ upper limits 
of $10^{8}$ -- $10^{9} M_{\sun}$ for the molecular gas mass within the host galaxies of the 
restarting radio galaxies.

Comparing our limits with those obtained in the comprehensive and consolidated CO observational 
study of radio galaxies of Evans et al. (2005), it is seen that for the lower value of 
the X-factor used, our limits are similar and the average upper limit to the molecular hydrogen mass 
we derive is lower.

What do our derived upper limits to the molecular gas content mean for the recurrence phenomenon in 
radio galaxies? The restarting radio galaxies all have upper limits that are lower than the molecular 
gas mass found in 3C\,293. If they possessed gas masses at a level of or above that found in 3C\,293 
they would have been detected. With 3C\,293, the only example of a restarting radio galaxy, setting the 
standard for fuel content in this class of sources, that of abundant molecular gas content, our 
observations have clarified that 3C\,293 may be atypical and that abundant molecular gas content of the 
level of $10^{9} M_{\sun}$ does not necessarily accompany the nuclear restarting phenomenon. 
Interestingly, 3C\,236, which is a re-starting source with a sub-galactic inner double source like 3C\,293,
as well as having abundant dust content is not detected in CO showing that molecular gas of the amount 
found in 3C\,293 is not a necessary condition for restarting of nuclear activity. 

None of the five normal, giant radio galaxies has a detection in CO. It appears that
the normal GRGs too in general do not possess molecular gas at the level of 
few $10^{9} M_{\sun}$ and that they are not hosts to unusually large stores of molecular 
gas in their nuclear regions. It is interesting to note that two of these large radio galaxies, 
0319$-$454 and 0511$-$305, show obvious signs of interactions -- suggesting
presence of large quantities of dust and accompanying gas (Subrahmanyan et al. 1996;
Saripalli et al. 1994; Bryant \& Hunstead 2000) and the two GRGs NGC~315 and NGC~6251 
are known to have nuclear dust disks.

Searches for molecular gas in radio galaxies with known strong IRAS detections 
(e.g. Evans et al. 2005) have shown that 
molecular gas is not commonly seen in these objects: less than one third are detected
in CO. The detections abound in compact and FR-I radio galaxies and tend to avoid powerful,
large FR-II radio galaxies.
In Section 1 we gave the rationale underlying the search for molecular gas in restarted
radio galaxies. We ask the question whether the interruption to the jet production is a result
of depletion of fuel necessary for central engine activity or an interruption to the 
fuel flow along the accretion disk, say, as a result of an interaction with a neighboring galaxy 
(e.g. Schoenmakers et al. 2000a; Kaiser et al. 2000). Based on the premise that all 
AGN activity requires fuel and
that the fuel is depleted in time, the molecular gas detection rate for the restarted 
radio galaxies is expected to be at least intermediate between the large FR-IIs and compact radio galaxies 
if the recurrence in activity is a result of a new accumulation of fuel. 

For the small size of our sample of nine radio galaxies, 
the formal CO detection rate for the restarted radio galaxies can be stated as
(11~$\pm~11)\%$. This value is  
small compared to that found for compact sources ($\sim40\%$; Evans et al. 2005). 
The lone detection among 
the large FR-IIs in Evans et al. (2005), 3C\,293, that distinguishes itself as the 
only known example of an FR-II radio galaxy possessing abundant molecular 
gas (Evans et al. 1999) is also a fine example of a restarting radio galaxy. Even excluding this
source from among the large FR-IIs in Evans et al. (2005) the statistically weak detection rate
for restarted radio galaxies only suggests that they are similar to the
large FR-IIs with respect to their molecular gas content: that abundant molecular gas presence
eludes them just like the FR-IIs. This could be viewed as supporting disruption to
the fuel flow (e.g. instabilities in the accretion disk, see Wada 2004) as a cause for the 
interruption 
to the jet production rather than a depletion of fuel. More sensitive upper limits
would need to be obtained however for a more robust comparison with the upper limits achieved 
for FR-II radio galaxies.

How do the limits compare with the estimates of fuel required for powering radio galaxies? 
With an Eddington accretion rate for a typical central black hole mass of 
 $10^{9}~{\rm M}_{\sun}$ of 5 to  $40~{\rm M}_{\sun}~{\rm yr}^{-1}$ (for maximally spinning and
non-rotating black holes, respectively), access to fuel masses
of at most $5 \cdot 10^8$ to $4 \cdot 10^9 M_{\sun}$ would be required for sustaining the activity over
intrinsic ages of $\sim 10^8$ yr (Hopkins et al. 2006). The upper limits we
derive are mostly of the order of few $10^8$ M$_{\sun}$. The observations suggest
prevalence of sub-Eddington accretion processes among the restarted radio
galaxies, a scenario that was recently suggested also for FR-I and FR-II
radio galaxies based on an analysis of black hole mass estimates and
nuclear luminosities (Marchesini et al. 2004).

We note that the targets in our sample of restarted radio galaxies 
were not selected on the basis of their IRAS flux densities unlike the sources in
Evans et al. (2005); only two are comparable in strength to Evans' sources. 
For three IRAS detected restarted radio galaxies in our sample, the IR luminosities 
are comparable to those of the radio galaxies in Evans' sample. However, differences 
in selecting the sources in the two samples (with respect to their IRAS fluxes) 
as well as the small numbers make a direct comparison difficult. For this reason we
cannot rule out a possible influence of pre-disposition to IRAS flux strengths 
on the molecular gas detection rates.

An interesting possibility is the existence of fuel in atomic form. This can be tested with studies 
of HI absorption against the compact cores of radio galaxies. Such studies
have been carried out (e.g. van Gorkom et al. 1989; Morganti et al. 2001) with the result that 
although only about one third are detected it is once again the compact radio galaxies 
that predominantly constitute the detections. However the studies have targeted strong radio core objects 
and a deeper study is needed that includes weak cores found in larger radio galaxies. We have commenced 
such a study with the Australia Telescope Compact Array of a large sample of radio galaxies with a range 
in linear sizes. The results will be presented in a future publication. 

Among the restarting radio galaxies in Table~3 only four have been included in samples of HI absorption 
studies. Three out of four (3C\,236, 3C\,293 and 4C\,29.30) have been detected in HI absorption against the cores 
indicating presence of atomic hydrogen gas close to the centres. Interestingly, the three also have among 
the most compact inner sources. Saikia et al. (2006) report detection of HI absorption against the core of 
a restarted radio galaxy, J1247+6723, that has a compact, 14~pc inner double. While there is evidence for
presence of atomic hydrogen in at least some restarting radio galaxies its role as fuel
awaits a systematic HI study of a larger sample made with high resolution
to resolve the radio structures.

\begin{table}
  \caption{Derived H$_2$-masses, assuming different values for the X-factor. Column (a) gives the 
lower limit with ${\rm X} = 3.7\cdot10^{23}\;{\rm m^{-2} (K\;km\;s^{-1})^{-1}}$, column (b) the upper 
limit with  ${\rm X} = 2.1\cdot10^{24}\;{\rm m^{-2} (K\;km\;s^{-1})^{-1}}$}
\begin{tabular}{|lcccc}
\hline
Source  & line        & \multicolumn{2}{c}{H$_2$  mass}          & Reference \\ 
        & transition  & \multicolumn{2}{c}{[$10^9$ $M_{\sun}$]} &  \\ 
        &             & (a)           & (b) &  \\
      
\hline
4C\,12.03 &CO(1-0) &$<1.12$  & $<6.35  $   & this work\\
          &CO(2-1) &$<1.37$  & $<7.79  $   & this work\\
4C\,29.30 &CO(1-0) &$<0.38$  & $<2.16  $   & 1\\
3C\,219   &CO(1-0) &$<0.95$  & $<5.42  $   & this work\\
3C\,236   &CO(1-0) &$<0.63$  & $<3.58  $   & this work\\
          &CO(2-1) &$<1.05$  & $<2.28  $   & this work\\
4C\,26.35 &CO(1-0) &$<0.37$  & $<2.09  $   & this work\\
          &CO(2-1) &$<0.22$  & $<1.27  $   & this work\\
3C\,293   &CO(1-0) &  4.32   &  24.56      & 1\\
3C\,388   &CO(1-0) &$<0.21$  & $<1.19  $   & this work\\
          &CO(2-1) &$<0.17$  & $<0.98  $   & this work\\
3C\,424   &CO(1-0) &$<0.66$  & $<3.74  $   & this work\\
          &CO(2-1) &$<0.42$  & $<6.40  $   & this work\\
3C\,445   &CO(1-0) &$<0.33$  & $<1.90  $   & this work\\
          &CO(2-1) &$<0.13$  & $<0.76  $   & this work\\
          & &         &             &           \\
          & &         &             &           \\
NGC\,315  &CO(1-0) &$<0.041$  & $<0.234$   & 2\\
          &CO(2-1) &$<0.013$  & $<0.072$   & 2\\
0319-454  &CO(1-0) &$<0.77$  & $<4.38  $   & this work\\
          &CO(2-1) &$<0.51$  & $<2.90  $   & this work\\
0503-286  &CO(1-0) &$<1.98$  & $<11.22 $   & this work\\
          &CO(2-1) &$<0.83$  & $<4.73 $    & this work\\
0511-305  &CO(1-0) &$<2.44$  & $<13.85 $   & this work\\
          &CO(2-1) &$<2.08$  & $<11.81 $   & this work\\
NGC\,6251 &CO(1-0) &$<3.85$  & $<21.83  $  & 3\\
\hline
\end{tabular}
References: 1: Evans et al. (2005), 2: Braine et al. (1997),
3: Elfhag et al. (1996)
\end{table}

\section{Summary}
We have presented results of a search program for molecular gas in a sample of powerful 
radio galaxies that are thought to be undergoing a second epoch of nuclear activity. The search
was carried out using the IRAM 30-m telescope on Pico Veleta 
in the CO J(1--0) and J(2--1) line transitions of the CO molecule, a known
effective tracer of molecular hydrogen. The motivation for the search for molecular
gas in these radio galaxies was to throw light on the possible causes of the
restarting of nuclear activity after a period of quiescence. One possible scenario is
depletion of a previous tranche 
of fuel followed by recent accumulation of fuel after a period of time and another possibility
is the continuation of the original fuelling process after an interruption. We expect that
the former possibility, where a recent fuel accumulation has restarted the nuclear
activity in these sources would increase their detectability in CO.

We discuss the CO detection rate for a representative sample
of nine restarting radio galaxies, seven of which were observed for the first time.
We report relatively sensitive upper limits that suggest there is no indication of
molecular gas mass larger than few $10^{8}$ -- $ 10^{9} M_{\sun}$. Abundant molecular
gas content therefore does not necessarily seem to accompany the nuclear restarting phenomenon. 
The restarting radio galaxies are found to be as deficient
in molecular gas as the edge brightened FR-II radio galaxies suggesting that there has been no recent
accumulation of molecular gas. The interruption 
to the jet production in these sources may have been a result of instabilities in the earlier fuel flow rather 
than due to a depletion of fuel. 
For comparison, we also discuss the molecular gas properties of five 
normal giant radio galaxies, three of which were observed by us using SEST.  These sources are
thought to form the parent sample of restarting sources. All five have only upper limits 
suggesting a lack of large ($\ga10^{9} M_{\sun}$) molecular gas content in these 
largest and among the oldest of radio galaxies.

\section*{Acknowledgments}
We are grateful to the referee whose comments have helped in improving the paper significantly.
KHM thanks the staff of the Paul Wild Observatory in Narrabri for their warm hospitality during 
a work stay. This work was supported by a Bilateral Science Agreement between CNR and CSIRO
(132.46.1/006297) and the Distinguished Visitor Programme of ATNF-CSIRO and has 
benefited from research funding from the European Union's Sixth Framework Programme under RadioNet R113CT2003 5058187.
Part of the observations were carried out with the IRAM 30-m telescope on Pico Veleta. IRAM
is supported by INSU/CNRS (France), MPG (Germany) and IGN (Spain). LS is grateful for the support 
provided by the staff at SEST during the observations.
This research has made use of the NASA/IPAC Extragalactic Database (NED) which is operated
by the Jet Propulsion Laboratory, California Institute of Technology, under contract with
the National Aeronautics and Space Administration.

\end{document}